\PassOptionsToPackage{warn}{textcomp}
\documentclass[sigconf]{acmart}
\usepackage{booktabs} % For formal tables
\usepackage{times}
\usepackage{epsfig}
\usepackage{graphicx}
\usepackage{amsmath}
\usepackage{amssymb}
\usepackage{amsmath,}
\usepackage{algorithm}
\usepackage[noend]{algpseudocode}
\usepackage{pseudocode}
\usepackage{multirow}
\usepackage{caption}
\usepackage{hhline}
\usepackage{xcolor}
\usepackage{dblfloatfix}
\usepackage{enumitem}
\usepackage{ulem}
\usepackage{color, colortbl}
\usepackage{sidecap}

\usepackage{bm}
\usepackage{multirow}
\usepackage{todonotes}
\usepackage{umoline}
\usepackage{xspace}
\usepackage{soul}
\usepackage{graphicx}
\usepackage{url}
\usepackage{subfigure}
\usepackage{amsmath,siunitx,booktabs,cuted,capt-of}
\usepackage{tabularx,ragged2e} % <-- new
\usepackage{rotating}
\usepackage{makecell, tabularx}

\sidecaptionvpos{figure}{c}

\setcopyright{none}
\acmDOI{10.24251/HICSS.2020.790}
\acmISBN{978-0-9981331-3-3}

\acmConference[HICSS '20]{Hawaii International Conference on System Science 2020}{January 7--10, 2020}{Maui, Hawaii}
\acmYear{2020}
\definecolor{lightblue}{rgb}{0.8,0.93,1}

\graphicspath{{figures/}} 

\newcolumntype{b}{>{\columncolor{lightblue}}c}

\begin{document}

\setlength{\belowdisplayskip}{3pt} \setlength{\belowdisplayshortskip}{3pt}
\setlength{\abovedisplayskip}{3pt} \setlength{\abovedisplayshortskip}{3pt}
\setlength{\textfloatsep}{1pt}

%\title{\sys{}: \underline{Gene}tic Evolution for Acquiring \underline{C}ompact \underline{AI}
\title{Digit Recognition From Wrist Movements and Security \\ Concerns with Smart Wrist Wearable IOT Devices}

% \titlenote{Produces the permission block, and
% \subtitle{Subtitle}
% \subtitlenote{The full version of the author's guide is available as
%   \texttt{acmart.pdf} document}

%%% The submitted version for review should be ANONYMOUS
% \author{Mojan Javaheripi,~~~Mohammad Samragh, ~~~Tara Javidi, ~~~Farinaz Koushanfar}
% \affiliation{%
%   \institution{ECE Department, UC San Diego, USA}
% }
% \email{mojan@ucsd.edu}
\author{Lambert T. Leong}
\affiliation{%
  \institution{CS Dept., University of Hawaii, USA}
}
\email{lambert3@hawaii.edu}
\author{Sean Wiere}
\affiliation{%
  \institution{MBBE Dept., University of Hawaii, USA}
}
\email{swiere@hawaii.edu}

\iffalse
\author{Tara Javidi}
\affiliation{%
  \institution{ECE Dept., UCSD, USA}
}
\email{tjavidi@ucsd.edu}
\author{Farinaz Koushanfar}
\affiliation{%
  \institution{ECE Dept., UCSD, USA}
}
\email{farinaz@ucsd.edu}
\fi

% \renewcommand{\shortauthors}{B. Trovato et al.}

\begin{abstract}
In this paper, we investigate a potential security vulnerability associated with
wrist wearable devices.  Hardware components on common wearable devices include
an accelerometer and gyroscope, among other sensors.  We demonstrate that
an accelerometer and gyroscope can pick up enough unique wrist movement information
to identify digits being written by a user.  With a data set of 400 writing
samples, of either the digit zero or the digit one, we constructed a machine
learning model to correctly identify the digit being written based on the
movements of the wrist.  Our model's performance on an unseen test set resulted
in an area under the receiver operating characteristic (AUROC) curve of 1.00.
Loading our model onto our fabricated device resulted in 100\% accuracy when
predicting ten writing samples in real-time.  The model's ability to correctly
identify all digits via wrist movement and orientation changes raises security
concerns.  Our results imply that nefarious individuals may be able to gain
sensitive digit based information such as social security, credit card, and
medical record numbers from wrist wearable devices.
\end{abstract}

\keywords{
\vspace{-0.1cm}Mahine Learning, Cyber Security, Wearables, IOT, Sensor, Artificial Intelligence}

\maketitle

\vspace{-0.3cm}

\section{Introduction}

Wearable smart technologies are becoming cheaper, more accessible, and thus more
common.  The wrist is an ideal location for wearable technologies and oftentimes this technology is in the form of a smart watch.  Smart watches afford
more functionality than just keeping track of the time and are often 
equipped with various hardware, such as infrared sensors, accelerometers,
gyroscopes, etc.  These various on-board hardware allows the user to track many
personal metrics that have implications for health and productivity benefits~\cite{wrist_review}.
While many features which take advantage of wrist wearable hardware output
already exist, the output variety is vast and all use cases have not yet been
explored.  Exploration into new ways to use wearable output metrics could result
in beneficial as well as malicious use cases.   
%In this work we sought to explore a new application for common on-board hardware of wrist wearable technologies.
%%% handwritting detection
In this work we investigate the potential of using wearable output metrics to
capture and predict hand written digits from users.  An individual's wrist
undergoes subtle movements and orientation changes~\cite{korean_vel_model,
ocd_hand} when writing different digits.  We hypothesized that these subtle
wrist movements and orientation changes are unique to the digit being written
and machine learning can be used to accurately classify the written digits. 

Hand written digit recognition from wrist movement and orientation has security
implications which include nefarious individuals gaining sensitive information
from users wearing smart wrist devices.  Sensitive information is often in
the form of digits such as social security, credit card, and medical record
numbers.  In addition, many wearable devices are connected to the internet and
recorded data is stored in the cloud.  Machine learning models which
can classify hand written digits from wrist movement and orientation could, in
theory, be feed data stored in the cloud to retroactively gain sensitive user
information.  

The remainder of this paper is organized as follows:
Section~\ref{sec:relatedwork} looks at previous work on machine learning and
handwriting recognition; Section~\ref{sec:methods} details our hardware design,
experimental design, and the construction and tuning of our machine learning
model; Section~\ref{sec:results} our model's performance is reported and  we
discuss our findings; Section~\ref{sec:conclusion} concludes this paper
with the implications of our findings in the scope of wrist wearable user security and
directions of future research.

\iffalse
Common wrist wearables including the
Apple Watch, Fitbit, and Samsung Galaxy Watch, contain the hardware needed to
capture movement and orientation~\cite{apple, samsung, fitbit}. 

Additionally, we believe that common wrist wearable hardware can capture these
subtle movements and machine learning can be used to classify written digits
based off particular wrist movements.  Any truth to our hypothesis can lead to
useful applications but it also raises security concerns.  However, security
becomes an issue when individuals write down sensitive information such as
passwords or social security numbers because wearable technology hardware may be
able to record this information.
\fi
%%% security, cloud storage, lots of training data

%%% The remainder of this paper ...

\section{Related Work} \label{sec:relatedwork}

Common wrist wearables, which include the Apple Watch, Fitbit, and Samsung
Galaxy Watch, contain the hardware capable of capturing movement and
orientation~\cite{apple, samsung, fitbit}. These hardware includes
accelerometers and gyroscopes, which have been shown to provide useful data
needed to identify fine motor task~\cite{Zeng_finemotor}.  In fact, other works
have shown that accelerometers mounted on the wrist have the sensitivity to
identify tremors associated with different neuro-muscular diseases such as
Parkinson's~\cite{wile_smartwatch_tremor} as well as
seizures~\cite{ml_wrist_seizure}.  Hand writing is a fine motor task and these
works lead us to believe that accelerometers and gyroscope hardware are
sufficient for measuring the movements the wrist undergoes.

Wearable internet of things (IoT) devices provide a constant data stream and
result in a considerable amount of data.  Machine and deep learning offers many
tools and techniques to analyze the vast and copious amounts of wearable IOT
data~\cite{iot_deep_opportunities}.  Various research efforts are aimed at
leveraging machine learning models to help make sense of all data and correlate
them to particular task and
activity related to sports performance, health, and safety~\cite{davila2017, ml_wrist_seizure}.%machine learning helps recog many task such as physical activity davila,  more specific to our work on the wrist and hand gestures from

In our work, we focus specifically on wrist wearables and machine learning
models built around the corresponding data.  Data from just a wrist mounted
gyroscope alone has been shown to be adequate for building a machine learning
model to detect hand gestures for a novel human computer interaction  (HCI)
device~\cite{ml_gest_recog}.  Several groups have explored machine learning to
build models that perform writing recognition task from wrist wearable device
output.  In one instance, additional custom sensors were placed on the upper
forearm and on the finger tip to capture additional information needed to
correctly classify hand gestures~\cite{fingerwritting}.  They were also able to
identify characters written with one's index finger with an accuracy of
95\%.  However, the strength of their model is likely attributed to the data
coming from the finger sensor more so, than the wrist sensors.  Our work aims
to perform written digit recognition from sensors placed solely on an individuals
wrist.  Word level recognition from smart watches was explored by Xia et
al.~\cite{motionhacker}. Their model was able to achieve an accuracy of
48.8\% on word level recognition based off wrist movement and they highlighted
potential security concerns of their results.  Letter level recognition with
smart watches was also explored~\cite{textmotion}.  In that work, writing tasks
were performed on whiteboards and  audio input from an on-board microphone was used
for segmentation, which helped recognition accuracy.  These works assured us that
sensor data from wrist wearable devices provides sufficient data for building machine
learning models to perform written recognition tasks.

Security issues associated with IoT devices is a popular and an ever growing area
of research.  IoT devices have been shown to be easily
compromised~\cite{wearable_security_attacks} and work has been done, using machine
learning, to help improve IoT security and detect threats on IoT
devices~\cite{Azmoodeh2018, ml_detect_attack}.  Our work is not necessarily concerned with the
issue of compromised wrist wearable IoT devices rather, it seeks to exploit a
nefarious use case for already available device data. An example of security
exploits on readily available wearable device data can be seen
in~\cite{bbc_news_2018, sly_2018}. In these articles, restricted areas such as
military bases have been mapped out just by having a wrist wearable user, with
security clearance, passively walk around secure areas. Wrist wearables
were not compromised in those instances but the use of the already available
data (e.g. GPS coordinates) posed an alarming security vulnerability.  Sensor
based attacks involving wrist wearables to capture keystrokes have been shown to
be possible~\cite{liu_keystroke}. In
another instance, Pandelea et al. showed that a machine learning model could be
built using data from on-board smart watch hardware to guess the password being
entered onto the device~\cite{password_guessing}.  In that paper, they showed
that pressing different keys on the smart watch corresponded to a different set of
movements and their model was able to map smart watch movements to different key
inputs.

\section{Methods}
\label{sec:methods}

For the purpose of our investigation we focus on handwritten digits.  More
specifically we focus only on the wrist movements associated with writing the
digit zero and the digit one. We formalize our problem, with respect to
machine learning, as a binary classification problem. Working with digits
mitigates issues that arise with the written differences in English upper and
lower case and cursive alphabet characters. 

\subsection{Hardware Design}

Accelerometers and gyroscopes are common amongst wrist wearables and these
hardwares are ideal for capturing the wrist movements and orientation. The
accelerometer can record the wrist acceleration in three planes, x, y, and z
and the gyroscope can capture the wrist angle or tilt during writing, in the x,
y, and z plane.  We fabricated our own devices equipped with an accelerometer
and gyroscope, similar to those seen in popular wrist wearables.  Designing our
own hardware allowed us to more accurately capture and label data for our
experiments.  

%Disengaging the switch would send a signal to output the
%data to a file and allow it to be labeled according to the digit that was
%written.

\iftrue
\begin{figure}[h!]
    \centering
    \subfigure[The ESP-32 feather board was used to handle processing and I/O from the LSM9D1
IMU]{\label{fig:esp}
    \includegraphics[width=.2\textwidth]{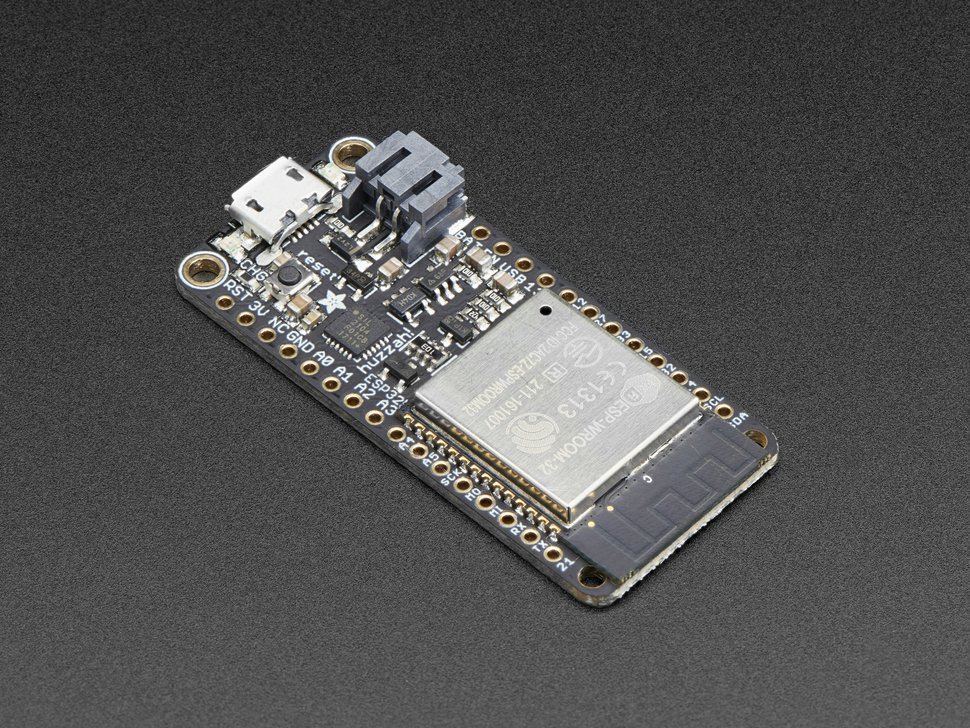}}
\hfill
    \subfigure[The LSM9D1 housed the accelerometer and gyroscope]{\label{fig:imu}
    \includegraphics[width=.2\textwidth]{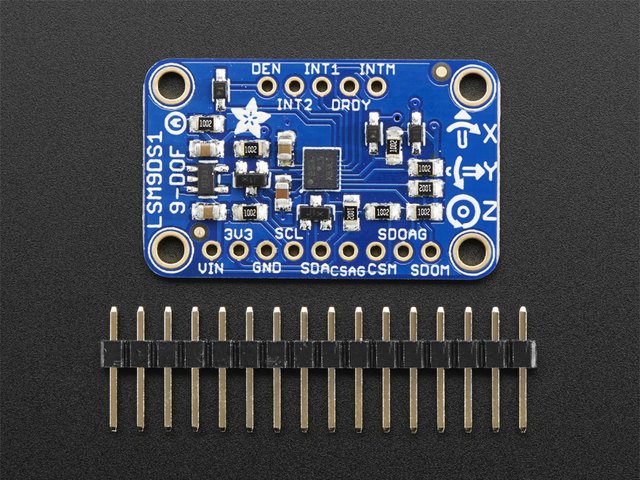}}

    \caption{Micro-controller and inertial measurement unit (IMU)}
    \label{fig:dist}
\end{figure}
\fi

We used Adafruit's ESP32 Feather~\cite{esp32r} micro-controller board and the
LSM9D1 inertial measurement unit (IMU)~\cite{lsm9d1r}.  The accelerometer and
gyroscope are housed on the IMU and serial peripheral interface (SPI) protocols
are used to communicate data recordings from the IMU to the ESP32 Feather board.
The IMU and micro-controller were connected and assembled into a single housing.
Acquiring labeled data or labeling data after collection can be expensive and
often requires a degree of processing and cleaning.  A switch was added
to the design of our device for the sole purpose of parsing and labeling data during
collection.  As a result, %idea being that 
the accelerometer and gyroscope data would only be recorded while the switch was
engaged and the switch was only engaged during the writing of either of the
digits. %recording would only take place when the switch was engaged.  
The switch allowed us to identify when writing began and ended within the
accelerometer and gyroscope data streams and also allowed for immediate labeling. 

\subsection{Data Collection}

We recruited participants who are right hand dominant and write with their right
hand.  The sensor housing was attached to the posterior side of the ulna and
radius at the most distal point from the body.  In other words, the housing was
attached to the top of the participants wrist, common in convention to how a
watch and wrist wearables are normally worn. Participants wrote out digits to
fill a 10 cm by 10 cm square region.  %While writing, participants were only allowed to make elbow contact with the writing surface.  
After the writing of each digit, the data was labeled with the appropriate digit
and saved to a file.  We collected a total of 400 writing samples of digits
which breaks down to 200 samples for the digit zero and 200 samples for the
digit one.

\subsection{Data Processing and Feature Engineering}

Seven data fields, as seen in Table~\ref{tab:og_feat}, were recorded from our
device during data collection. These fields include time, acceleration in three
planes (x, y, z), and pitch angle in three planes (x, y, z).  It is often
the case that certain digits can be written in different amounts of time.  For
instance, writing the digit one usually takes less time to write than the digit
zero.  To deter our machine learning algorithm from only learning on the time
it takes to write a digit, we extracted features which are uncoupled from time.
The features we extracted were the minimum, maximum, and mean for acceleration
and pitch angle in all three planes (x, y, z).

Using the acceleration features we were able to engineer a velocity and a
displacement feature.  Integrating over the acceleration yielded the velocity
and subsequently, integrating over the velocity yielded the displacement.
Velocities and displacements were calculated in all three planes.  The minimum,
maximum, and mean velocities for all three planes were added to our current
feature list. The total displacement in all three (x,y,z) planes was also added
to the feature list. Lastly, we calculated the total overall displacement and
added that to the new feature list.  As a result we transformed our original
seven features into 31 new features, shown in Table~\ref{tab:all_feat}, that are irrespective of time.

\begin{table}[h!]
\centering
\caption{List of original features gathered from the device}
\resizebox{.5\textwidth}{!}{
\begin{tabular}{lccr}
%\\
\toprule
%\multicolumn{4}{c}{{Original Features From the Device}} \\ \midrule
Metrics & Axis & Value Type& Feature Count \\ \midrule \midrule
Acceleration & x,y,z &  & 3 \\
Pitch Angle & x,y,z &  & 3 \\
Time &  & Total & 1 \\ \midrule
\textbf{Total Features} & \multicolumn{1}{l}{} & \multicolumn{1}{l}{} &
\textbf{7} \\ \bottomrule
\end{tabular}
}
\label{tab:og_feat}
\end{table}

\iftrue
%\begin{table}[t]  % will be placed at top of a page
\begin{table*}  % will be placed at top of a page
\caption{List of engineered features. Features used in final model}
%\begin{tabularx}{\linewidth}{ l CCr }
\begin{tabular}{ l ccr }
%\\
\toprule
%\multicolumn{4}{c}{{Engineered Features}} \\ \midrule
 Metrics & Axis &  Value Type& Feature Count \\ \midrule \midrule
 Acceleration & x,y,z & Minimum, Mean, Maximum & 9 \\
 Pitch Angle & x,y,z & Minimum, Mean, Maximum & 9 \\
 Velocity & x,y,z & Minimum, Mean, Maximum & 9 \\
 Displacement & x,y,z & Total & 4 \\ \midrule
\textbf{Total Features} & \multicolumn{1}{l}{} & \multicolumn{1}{l}{} &
\textbf{31} \\ \midrule
\end{tabular}
\label{tab:all_feat}
\end{table*}
\fi

\subsection{Principal Components Analysis and Class Separability Investigation}
\label{sec:methods_pca}

Principal components analysis (PCA) was used to evaluate the explained variance
of each of the 31 re-engineered features. Data was normalized and Scikit-Learns~\cite{sklearn} PCA
module was used to perform PCA. It was found that the top three principal components explain
greater than 99.99\% of the variance. Mapping the principal components (PC) back to
the original features revealed that the top three components are the maximum
pitch angle in the z plane, the displacement in the x plane and the mean pitch
angle in the z plane, respectively.  Distributions for each of the three features were
generated with respect to class and are presented in Figure~\ref{fig:dist}.

\iftrue
\begin{figure}[h!]
    \centering
    \subfigure[Distribution of maximum pitch angle in the z plane with respect
to class.  gz = gyroscope z plane]{\label{fig:max_gz}
    \includegraphics[width=.35\textwidth]{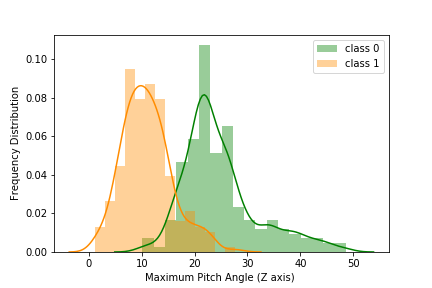}}
    \subfigure[Distribution of displacement in the x plane with respect to
class dx = displacement in x plane]{\label{fig:dx}
    \includegraphics[width=.35\textwidth]{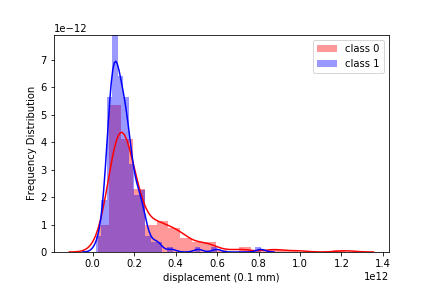}}
    \subfigure[Distribution of mean pitch angle in the z plane with respect to
class.]{\label{fig:mean_gz}
    \includegraphics[width=.35\textwidth]{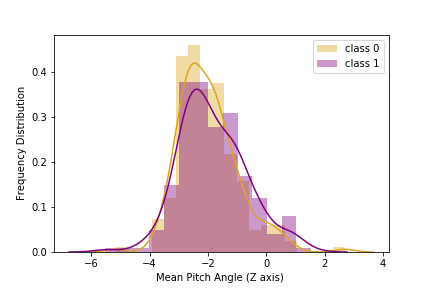}}

    \caption{Distribution of top three principal components with respect to
class}
    \label{fig:dist}
\end{figure}
\fi

Figure~\ref{fig:max_gz} shows the best separation of classes when compared to
the x displacement and mean z pitch angle.  There is still some, non-negligible
overlap in Figure~\ref{fig:max_gz} and other features may be needed to get clear
class separation. Figure~\ref{fig:dx} and~\ref{fig:mean_gz} have a considerable
overlap but some separation can be seen and including these features may help
define a better decision boundary. Scatter plots were generated to
investigate the separability of the two classes and a potential decision
boundaries.  Plots are shown in Figure~\ref{fig:scatter}.  %Figures are shown below.

\iftrue
\begin{figure}[h!]
    \centering
    \subfigure[x displacement versus maximum z pitch angle colored by
class]{\label{fig:dx_max_gz}
    \includegraphics[width=.35\textwidth]{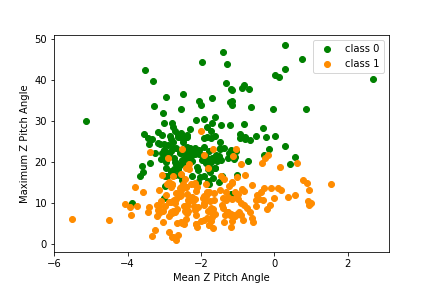}}
    \subfigure[mean z pitch angle versus maximum z pitch angle colored by
class]{\label{fig:dx_mean_gz}
    \includegraphics[width=.35\textwidth]{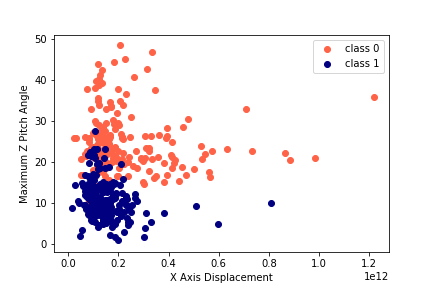}}
    \subfigure[x displacement versus maximum z pitch angle versus mean z pitch
angle colored by class]{\label{fig:3d}
    \includegraphics[width=.35\textwidth]{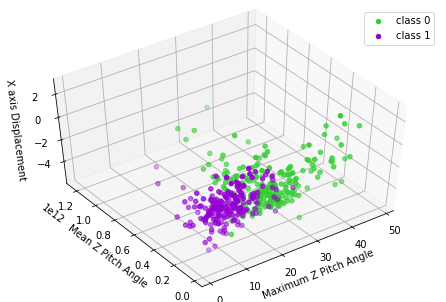}}
    \caption{Plots of top three principal components versus each other to look
for separability}
    \label{fig:scatter}
\end{figure}
\fi

Figure~\ref{fig:dx_max_gz} shows fairly good separation amongst the classes
however there there is some overlap in the middle where the two classes meet.
Figure~\ref{fig:dx_mean_gz} also presents with fairly good separation but there
is also overlap in the middle where the two classes meet.  Good separability can
be seen with respect to two of the top three principal components.
Figure~\ref{fig:3d} plots all three principal components against each other to
see if better separability can be seen in a higher dimension.  Overlap of the
two classes is still present, however, there is more separation seen in three
dimensions than seen in two dimensions.  This is somewhat expected as
greater separability is often observed in higher dimensions and using more
features may create more separation between the two classes. Therefore, it is
likely that more features are needed to generate a stronger model and thus we
chose to build a model utilizing all 31 features rather than just the top three
PC.
 
\subsection{Model and Hyper Parameter Tuning}

The dataset of 400 writing samples was randomly split into a training,
validation, and testing sets via a 60\%, 20\%, 20\% split.  The number of samples
for each class is shown in Table~\ref{tab:data_count}.  Our relatively modest
sample size led us to first explore simpler models rather than a deep learning
approach.  We explored ensembling methods which led us to use extreme gradient
boosting with the help of Sci-Kit Learn's xgboost package.  To evaluate our
model choice we set the following hyper parameters to the package defaults
as follows: number of boosting stages at 100, learning rate of 0.1,
max tree depth of six, and the auto tree algorithm.  The classifier was
trained on the training set with all 31 feature and the validation set achieved
an AUROC of 88.03\%.  Preliminary performance was good and we proceeded with
gradient boosting as our model choice.

\begin{table}[h!]
\centering
\caption{Breakdown of sample per class}
\resizebox{.3\textwidth}{!}{
\begin{tabular}{lrr}
\toprule
	Dataset & Class digit 0 & Class digit 1\\ \midrule \midrule
	Train & 122 & 118\\ \midrule
	Validation & 42 & 38\\ \midrule
	Test & 36 & 44\\ \midrule
	\textbf{Total} & \textbf{200} & \textbf{200} \\ \bottomrule
\end{tabular}
}
\label{tab:data_count}
\end{table}

\iffalse
% present AUROC of pca reduced val and test
Model performance is evaluated using the Area Under the Receiver Operating
Characteristic (AUROC). Using the same model and hyper parameters we trained on
the PCA transformed data set with the top three principal components. We
achieved an AUROC of 84.32\% on the PCA reduced validation set. PCA led to a
lower AUROC when compared to the full feature data set. As a result we
proceeded with the full 31 features and tuned model parameters.
\fi

% hyper parameter tuning with grid search cv
% 5 cv  and parameter ranges
Hyper parameters, which include the number of estimators, the learning rate,
the maximum depth of a tree, and the tree construction algorithm, were
optimized using an exhaustive grid search. Models were trained on different
combinations of hyper parameters using five fold cross-validation. A list of
hyper parameters and their explored ranges are presented in
Table~\ref{tab:params}.

\begin{table}[h!]
\centering
\caption{Hyper parameters and ranges explored}
\resizebox{.38\textwidth}{!}{
\begin{tabular}{lr}
\toprule
Parameter & Values and ranges\\ \midrule \midrule
n\_estimators & 1000, 2000, 3000, 4000, 5000 \\ \midrule
tree\_algorithm & hist, exact \\ \midrule
max\_depth & 1, 2, 3, 4, 5, 6, 7, 8 \\ \midrule
learning rate& 0.1, 0.3, 0.5 \\ \bottomrule
\end{tabular}
}
\label{tab:params}
\end{table}

The best hyper parameters, from the exhaustive grid search, which yielded the
best AUROC's on the validation, set are shown in Table~\ref{tab:best_params}.
The final models were retrained on the combined, train and validation, dataset
using five fold cross-validation.  The best hyper parameters, from
Table~\ref{tab:best_params} were used in the retraining of the final PCA model and full 31
feature model. 

% report best parameters

\begin{table}[h!]
\centering
\caption{Best hyper parameters used to train final models}
\resizebox{.25\textwidth}{!}{
\begin{tabular}{lr}
\toprule
Parameter & Values \\ \midrule\midrule
n\_estimators & 1000 \\ \midrule
tree\_algorithm & hist \\ \midrule
max\_depth & 1 \\ \midrule
learning rate& 0.1\\ \bottomrule
\end{tabular}
}
\label{tab:best_params}
\end{table}

\subsection{Performance Evaluation}

Model performance was mainly evaluated using the area under the receiver operating
characteristic (AUROC) curve.  AUROC values are reported as values between zero
and one where values closer to one indicate better performance.  The test set,
which is 20\% of the dataset that was never seen by the models, was used to
calculate the final AUROC values. Final AUROC values from both the final PCA
model and the final full feature model were compared to each other to evaluate
model performance.

\section{Performance Results}
\label{sec:results}

Two models were constructed using a different number of features.  The first model
was constructed using the top three principal components (PC) and the second model
used all 31 features.  The better of the two models was used to construct our
final model which was ported to our device for real time evaluations.

\subsection{PCA Model Results}

As mentioned in Section~\ref{sec:methods_pca}, the top three principal
components explain 99.99\% of the variance and we investigated if these
features were sufficient to build a good classifier.  The test set, which was
held out and not seen by the model during construction and hyper parameter
tuning, was offered to the model constructed with the three features, shown in
Table~\ref{tab:pca_feat}, corresponding to the top three PC.  The receiver
operating characteristic (ROC) curve and AUROC value is shown in
Figure~\ref{fig:pca_roc}.

\begin{table}[h!]
\centering
\caption{List of PCA reduced features}
\resizebox{.5\textwidth}{!}{
\begin{tabular}{lccr}
%\\
\toprule
Metrics & Axis & Value Type & Feature Count \\ \midrule \midrule
Pitch Angle & z & Maximum & 1 \\
Displacement & x & Total & 1 \\
Pitch Angle & z & Mean & 1 \\ \midrule
\textbf{Total Features} & \multicolumn{1}{l}{} & \multicolumn{1}{l}{} &
\textbf{3} \\ \bottomrule
\end{tabular}
}
\label{tab:pca_feat}
\end{table}

\iftrue
\begin{figure}[h!]
    \centering
    \subfigure[]{\label{fig:pca_roc}
    \includegraphics[width=.5\textwidth]{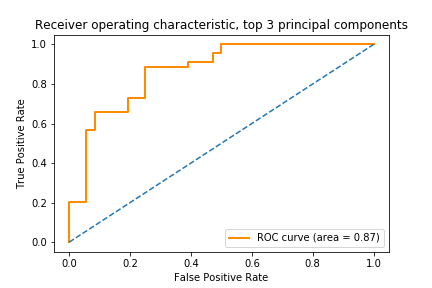}}
    \caption{ROC curve and AUROC values calculated as a result of running the test set through the model
trained with only the top three principal components. The dash line indicates random guess performance.}
    \label{fig:pca_roc}
\end{figure}
\fi

\iffalse
\begin{table}[h!]
\centering
\caption{Confusion matrix of PCA models performance on held out test set}
\resizebox{.4\textwidth}{!}{
\begin{tabular}{c >{\bfseries}r @{\hspace{0.7em}}c @{\hspace{0.4em}}c @{\hspace{0.7em}}l}
  \multirow{10}{*}{\rotatebox{90}{\parbox{1.1cm}{\bfseries\centering Actual\\ Digit}}} & 
    & \multicolumn{2}{c}{\bfseries Predicted Digit} & \\
  & & \bfseries 0 & \bfseries 1 &\\% \bfseries total \\
  & 0 & \MyBox{True}{Positive\\ 33} & \MyBox{False}{Negative\\ 8} &  \\[2.4em]
  & 1 & \MyBox{False}{Positive\\ 8} & \MyBox{True}{Negative\\ 36} &  \\
  %& total & P & N &
\end{tabular}
}
\label{tab:pca_cm}
\end{table}
\fi

\iftrue %TODO: true
\begin{table}[h!]
\setlength\extrarowheight{2pt}
\centering
\caption{Confusion matrix of PCA models performance on held out test set}
\resizebox{.33\textwidth}{!}{
\begin{tabular}{@{}cc|cc@{}}
\multicolumn{1}{c}{} &\multicolumn{1}{c}{} &\multicolumn{2}{c}{Predicted } \\ 
\multicolumn{1}{c}{} & 
\multicolumn{1}{c|}{} & 
\multicolumn{1}{c}{Digit 0} & 
\multicolumn{1}{c}{Digit 1} \\ 
\cline{2-4}
%\multirow[c]{2}{*}[-2.0ex]{\rotcell{Actual}}
\multirow{2}{*}{\rotatebox[origin=tr]{90}{Actual }}
& Digit 0  & 33 & 3   \\[1.5ex]
& Digit 1  & 8   & 36 \\ 
\cline{2-4}
\end{tabular}
}
\label{tab:pca_cm}
\end{table}
\fi

PCA was explored as a means of possibly reducing the dimensionality of the
data.  We were interested to see if any subset of the 31 features could be used
to build a strong model. Training a model with just the top three PC resulted
in an AUROC of 0.87, as seen in Figure~\ref{fig:pca_roc}.  Class break downs and
predictions by class are shown in the confusion matrix in
Table~\ref{tab:pca_cm}. Results indicate that
three feature (maximum z pitch angle, total x displacement, and mean z pitch
angle) may not contain enough information to define a clear decision boundary.
Dimensionality reduction offered the potential to build our model off of only
one sensor input which could have made our methods more applicable to more
devices.  For instance, if a model built on only gyroscope input could
correctly classify the written digits then accelerometer hardware would not be
needed.  Our PCA model suggested the contrary and more features were needed thus
both sensors are needed as well. 

\subsection{Full Feature Model Results} 

We constructed a second model in an attempt to achieve better performance than the three
PC model.  We used all 31 engineered features, shown in Table~\ref{tab:all_feat}, to train
a second model and the performance results are shown in Figure~\ref{fig:final_roc}.

\iftrue
\begin{figure}[h!]
    \centering
    \subfigure[]{\label{fig:final_roc}
    \includegraphics[width=.5\textwidth]{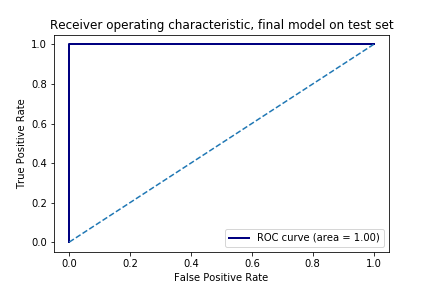}}

    \caption{ROC curve and AUROC values calculated as a result of running the test set through the model
trained on all 31 feature. The dash line indicates random guess performance.}
    \label{fig:final_roc}
\end{figure}
\fi

\iftrue %TODO: true
\begin{table}[h!]
\setlength\extrarowheight{2pt}
\centering
\caption{Confusion matrix of final models performance on held out test set}
\resizebox{.33\textwidth}{!}{
\begin{tabular}{@{}cc|cc@{}}
\multicolumn{1}{c}{} &\multicolumn{1}{c}{} &\multicolumn{2}{c}{Predicted } \\ 
\multicolumn{1}{c}{} & 
\multicolumn{1}{c|}{} & 
\multicolumn{1}{c}{Digit 0} & 
\multicolumn{1}{c}{Digit 1} \\ 
\cline{2-4}
%\multirow[c]{2}{*}[-2.0ex]{\rotcell{Actual}}
\multirow{2}{*}{\rotatebox[origin=tr]{90}{Actual }}
& Digit 0  & 36 & 0   \\[1.0ex]
& Digit 1  & 0   & 44 \\ 
\cline{2-4}
\end{tabular}
}
\label{tab:final_cm}
\end{table}
\fi

Figure~\ref{fig:final_roc} shows an improvement in model performance when all 31
features are used to train our model. The second model achieved an AUROC of
1.00 on our test set. This indicates that it was able to predict which digit
was written without any errors. Class break downs and full feature model
predictions by class are shown in the confusion matrix in
Table~\ref{tab:final_cm}. Although the top three PC explain the great
majority of the variance, the other components contain the information needed to
create a good decision boundary. Separability is often easier to observe in
higher dimensions and this seemed to be the case for our classification problem.
An AUROC value of 1.00 does not suggest the need for more feature engineering
and suggests that our current features are sufficient.  We are confident in our
model's generalizability due to its good performance on the test set and have
no reason to suspect any significant overfitting.

\iffalse
Using the best hyper parameters, presented in Table~\ref{tab:best_params}, we
trained using the training set and assessed the models performance on the
validation set.  The validation AUROC was 99.97\% which made us confident in our
model and hyper parameters.
%Some of the hyper parameters were manually manipulated, dispite the grid search
%results, to investigate how differece in certian parameters affect the model.  
As a final check, we ran the test set through the model and compared it's AUROC
to the validation
sets AUROC to evaluate the generalizability.  The test sets AUROC was 99.81\%
which slightly lower than the validation set and may be indicative of minor
overfitting.  Although there is minor overfitting, we believe we have a good
model with the best hyper parameters.  The final model is retrained on the full
data set with the best hyper parameters.
\fi

\subsection{Real Time Performance}
The second, full feature, model was retrained on the entire dataset with the
same optimum hyper parameters to produce our final model.  The final model was
loaded onto our hardware device so that real time predictions could be made. A
sample set of ten written digits, which breaks down to five samples of
written digit zero and five samples of written digit one, was evaluated in
real time.  Volunteers
were randomly assigned to write either the digit 0 or the digit 1. With
the device attached to their wrist, the volunteer wrote their assigned
digit and the devices output a prediction. Results from real time
testing are reported in Table~\ref{tab:real_cm}.
%Writers wrote zeros and ones in a random order and the model output a
	%prediction
%after each digit was written.  

\iftrue %TODO: true
\begin{table}[h!]
\setlength\extrarowheight{2pt}
\centering
\caption{Confusion matrix of final models performance during real time testing}
\resizebox{.33\textwidth}{!}{

\begin{tabular}{@{}cc|cc@{}}
\multicolumn{1}{c}{} &\multicolumn{1}{c}{} &\multicolumn{2}{c}{Predicted } \\ 
\multicolumn{1}{c}{} & 
\multicolumn{1}{c|}{} & 
\multicolumn{1}{c}{Digit 0} & 
\multicolumn{1}{c}{Digit 1} \\ 
\cline{2-4}
%\multirow[c]{2}{*}[-2.0ex]{\rotcell{Actual}}
\multirow{2}{*}{\rotatebox[origin=tr]{90}{Actual }}
& Digit 0  & 5 & 0   \\[1.5ex]
& Digit 1  & 0   & 5 \\ 
\cline{2-4}
\end{tabular}
}
\label{tab:real_cm}
\end{table}
\fi

\begin{table*}  % will be placed at top of a page
\caption{Performance metrics for the PCA, full feature models, and real time full feature models}
%\begin{tabularx}{\linewidth}{ l CCCC CCC @{}}
\resizebox{1.\textwidth}{!}{

\begin{tabular}{ l cccc ccc @{}}
%\multicolumn{1}{c}{} &\multicolumn{1}{c}{} &\multicolumn{2}{c}{Digit 0 }\\% &\multicolumn{2}{c}{Digit 1 } \\
%\begin{tabularx}{\linewidth}{@{} l CCCC @{}}
\toprule
Model & Accuracy (\%)  & Digit 0 Precision (\%) &  Digit 0 Recall (\%) & Digit 0 F1 Score (\%) & Digit 1 Precision (\%) &  Digit 1 Recall (\%) & Digit 1 F1 Score (\%) \\ \midrule
%Compound & Theoretical Yield (g) & Experimental Yield (g) & Percent Yield & Percent Composition of Total Product \\
\midrule
%Model & Acuraccy (\%) & Percision (\%) & Recall (\%) & F1 Score (\%) \\ \hline
%PCA & 85.00 & 93.35 & 79.55 & 85.90 \\ \midrule
	PCA & 86.25 & 80.49 & 91.67 & 85.74 & 92.31 & 81.82 & 86.75 \\ \midrule
	Full Feature & 100.00 & 100.00 & 100.00 & 100.00 & 100.00 & 100.00 & 100.00 \\ \midrule%\hline
	Real Time, Full Feature & 100.00 & 100.00 & 100.00 & 100.00 & 100.00 & 100.00 & 100.00 \\ %\hline
%Total Final Product & 2.90 & 1.6853 & 58.1 \\[1ex]
%2-methyl-2-butene   & 2.60 & 1.56   & &   92.60756 \\
%2-methyl-1-butene   & 0.30 & 0.12   & &    7.39244 \\
\bottomrule
\end{tabular}
}
\label{tab:model_met}
\end{table*}

As seen from the confusion matrix, in Table~\ref{tab:real_cm}, the final
model was able to predict all written digits accurately which results in an
AUROC of 1. 
%Our final model performed with 100\% accuracy, precision, recall, and F1 score as seen in Table~\ref{tab:model_met}.  
Real time performance results reassures us of our models generalizability and,
again, does not lead us to suspect significant overfitting.  Performance
metrics, including accuracy, percision, recall and F1 score, are reported for
the PCA, full feature, and real time full feature models in
Table~\ref{tab:model_met}.

% present AUROC of validation and test
% talk about what it all means
% re gen model with all data points

\section{Conclusions}
\label{sec:conclusion}
%can capture subitle movements with hardware
%  hardware is common in wrist wearables
%subtile movemvents are unique and machine learning can predict well
%potential security issues
%  private written information can be recorded or stolen
%
Our findings imply a potential security vulnerability that is associated with
wrist wearable devices. Accelerometers and gyroscopes, which we used, are common
hardware on-board wrist wearables. We demonstrate the ability to capture the
subtle movements and position changes of the wrist with those hardware during
writing. Using machine learning, we were able to identify that the wrist
movements involved in writing the digit zero is unique and different to the
wrist movements involved in writing the digit one. As a result, a robust machine learning
model was constructed which demonstrated perfect real-time
prediction performance. Our results imply a plausible reality where sensitive
information can be recorded from users during writing while wearing some smart
watch or wrist wearable device.  In addition, our potential security exploit results from
using already available data from smart wrist wearables.   Our methods do not
involve nor require compromising the wearable devices themselves.   While the
machine learning model we developed is simple and only for the binary
classification of two written digits, it is an important first step and brings
awareness to some security vulnerabilities associated with wrist wearables.

\section{Future Work}

We hope to explore how our data and model relates to left hand
dominant users. It is hypothesized that since your left hand is a mirror
image of the right that simply flipping signs or the direction vector will
lead to the correct solution.  More data, specific to left handed users, is needed
to explore how handedness affects generalizability of machine learning writing
recognition models. 

The size of our data set was modest and only contained wrist movement data
for the digits one and zero. Our binary classification problem was not hard to
solve and thus a simpler model was sufficient which also raises concerns. The
use of a simpler machine learning model implies that capturing users private
information may be a trivial task.  Our current findings warrant further work
to aggregate wrist movement data for the writing of all ten digits, zero
through nine. Working with all ten digits presents as a multi-class classification
problem and while it may be more difficult there exists more powerful tools
which were not explored in our work.  In fact, deep neural networks could
potentially handle classifying wrist movements for all ten digits fairly
easily. More work is needed to explore these concepts to improve and maintain
security for the vast variety of wearable IoT devices.

\bibliographystyle{ACM-Reference-Format}
\bibliography{bib} 
%\bibliography{sample-bibliography} 

% \input{tables}

\end{document}